\documentclass[twocolumn,preprintnumbers,amssymb,amsmath,superscriptaddress,letterpaper]{revtex4}[12pt]

\usepackage{graphicx}
\usepackage{dcolumn}
\usepackage{bm}
\usepackage{natbib}
\usepackage{pstricks}
\usepackage{epsfig}
\usepackage{epstopdf}

\newcommand{\be}{\begin{equation}}
\newcommand{\ee}{\end{equation}}
\newcommand{\bea}{\begin{eqnarray}}
\newcommand{\eea}{\end{eqnarray}}

\begin{document}

\title{The Cosmological Constant Potential:\\ a resolution to the Hubble tension via the cosmological sound horizon}

\author{Nima Khosravi}
\email{nima@sharif.edu}
\affiliation{Department of Physics, Sharif University of Technology, Tehran 11155-9161, Iran}
\affiliation{Department of Physics, Shahid Beheshti University, 1983969411, Tehran, Iran}

\date{\today}

\begin{abstract}
The cosmological constant term can be seen as a constant potential for a (scalar) field. In this viewpoint, at late times, the field is stopped rolling and behaves as a cosmological constant ($w=-1$). While at the early universe, its kinetic term can be dominant and behaves as a stiff fluid ($w=+1$). This new phase lowers the cosmological sound horizon  by increasing the Hubble parameter for very high redshifts. Consequently, the lower cosmological sound horizon results in the higher Hubble constant at the present time. This early phase ends before the photon decoupling, so we do not expect any new physics after the last scattering surface. We checked this model in the presence of (reduced) CMB, BAO's and $H_0$ datasets and could show the Hubble tension is fully relieved.

\end{abstract}
\maketitle
\section{Introduction:}
The tensions and anomalies can lead us to understand better and deeper our theoretical models as well as observations. In recent years the Hubble ($H_0$) tension in cosmology made us to think more about the standard model of cosmology $\Lambda$CDM which includes the cosmological constant ($\Lambda$) and the cold dark matter (CDM) as main ingredients. The $H_0$ tension shows a discrepancy between the local direct measurements of $H_0$ by distance ladders \cite{R16,Riess:2019cxk,Riess:2021jrx} and the inferred value by  the cosmic microwave background (CMB) \cite{Aghanim:2018eyx}. Although, the CMB is an indirect measurement which means one needs to assume the $\Lambda$CDM to connect physics of early universe to the present time observations. Though $H_0$ tension seems to be the most severe one but there are many other reported tensions and anomalies: the $S_8$ tension \cite{Heymans:2020gsg}, the lensing anomaly  \cite{Aghanim:2018eyx} and the spatial anomalies \cite{Schwarz:2015cma,Planck:2019evm}. However our main focus will be on the $H_0$ tension in this work.

If these tensions are not statistical flukes or systematics then there is a request to explain them by new theoretical models, the models beyond the standard $\Lambda$CDM \cite{DiValentino:2021izs}. There are a vast literature trying to address the $H_0$ tension by modifying different parts of the standard model. The natural scenarios seem those which are related to the late time dark energy  \cite{DiValentino:2021izs}. But there are some counterarguments against them \cite{Efstathiou:2021ocp}. A way to see the problem is to look at BAO's: BAO's share the same physics with CMB. This makes solving the $H_0$ tension by the late time modifications hard since it seems the late time modifications affect CMB and BAO differently. Then (based on this argument) there are early modification of cosmological constitutes \cite{Kamionkowski:2022pkx,Knox:2019rjx,Poulin:2018cxd}. The idea is to modify the sound horizon scale, $r_s$, and since the CMB put constraint on a combination of $H_0$ and $r_s$ then one can be able to address the Hubble tension indirectly. Again there is still some arguments against these models \cite{Vagnozzi:2023nrq}. 

In this work, we study (maybe) the simplest possible model i.e. the standard $\Lambda$CDM but we change our viewpoint on the cosmological constant term. We look at this constant as a constant potential where its associated (scalar) field stoped rolling at the late times. The early dynamics of this field can affect the cosmological expansion and consequently $r_s$ and $H_0$ as we will show. 

\section{Model:}
The model is (almost) exactly the standard $\Lambda$CDM model with a twist of interpretation on the cosmological constant, $\Lambda$, which opens a new way to look at. We assume the $\Lambda$ term is a potential of a (scalar) field i.e. $V(\phi)=\Lambda$. This means we have a very specific case of quintessence model 
\begin{eqnarray}
{\cal{L}}=\sqrt{-g}\,\bigg[ R - \partial_\mu\phi\,\partial^\mu\phi\,-\,V(\phi)\bigg]
\end{eqnarray}
where as mentioned above $V(\phi)=\Lambda$ is a constant. Now the equation of motion for the scalar field will be
\begin{eqnarray}
\Box\phi+V_{,\phi}=0
\end{eqnarray}
which for our scenario reduces to $\Box\phi=0$ with two solutions (in the case of isotropic and homogeneous cosmological background)
\begin{eqnarray}
\phi &=& \text{constant},\\
\phi &\propto& a^{-3}
\end{eqnarray}
where $a$ is the scale factor for the FRW metric. The first solution gives exactly the standard $\Lambda$CDM model but the second one has some non-trivial and interesting properties. However since the equation in linear in $\phi$ so the superposition of both is a solution. To go further let's switch to the perfect fluid prescription for the scalar fields
\begin{eqnarray}
\rho_\phi &=&  \partial_\mu\phi\,\partial^\mu\phi\,+\,V(\phi)\doteq  \partial_\mu\phi\,\partial^\mu\phi\,+\Lambda,\\
p_\phi &=&  \partial_\mu\phi\,\partial^\mu\phi\,-\,V(\phi)\doteq  \partial_\mu\phi\,\partial^\mu\phi\,-\Lambda
\end{eqnarray}
where $\doteq$ means in our scenario. So for the case of background cosmology we will have
\begin{eqnarray}
\rho_\phi &=&\alpha\,a^{-6}+\Lambda,\\
p_\phi &=&\alpha\,a^{-6}-\Lambda
 \end{eqnarray}
where $\alpha$ is an integration constant. The above relation says for very large $a$'s we have the cosmological constant. The $\alpha$ term is representing non-vanishing initial velocity of the scalar field $\dot{\phi}\neq0$. This non-vanishing initial velocity term is diluted very rapidly by $a^{-6}$. The corresponding equation of state, $w_{CCPot}$, is given by
\begin{eqnarray}\label{wCCPot}
w_{CCPot}=\frac{\alpha\,a^{-6}-\Lambda}{\alpha\,a^{-6}+\Lambda}=\frac{e^{-\beta}-a^{6}}{e^{-\beta}+a^{6}},
 \end{eqnarray}
 where $e^{-\beta}\equiv\alpha/\Lambda$. This means we are dealing with a fluid which transits from a stiff matter era ($w=+1$) at the very early times ($a\rightarrow 0$) to the cosmological constant ($w=-1$) for the late times ($a\rightarrow 1$). The equation of state of the stiff matter, $w=+1$, is at the border of satisfying the causality. For $w>1$ the speed of sound exceeds the speed of light which is not acceptable though there are some arguments \cite{Ellis:2007ic}. The cosmology of stiff matter is studied in \cite{Chavanis:2014lra} for a specific scenario and earlier in \cite{Zeldovich:1961sbr}. The same behavior can also happen in the ultra slow roll inflation \cite{Dimopoulos:2017ged} where we will make a comment on it later.

 \section{The Hubble tension and/or the $r_s$ tension}
 The Hubble tension states the local direct measurement of the Hubble constant is not consistent with the derived one from the CMB measurements. In this direction, we would like to focus on the derived $H_0$ value from the CMB. The CMB cannot directly constrain ``$H_0$" but its constraint is on ``$r_s\times H_0$"  \cite{Knox:2019rjx}. It is the reason that the $r_s$ tension is also studied directly \cite{Aylor:2018drw}.
 
 So this degeneracy in the CMB measurement suggests a solution for the $H_0$ tension: instead of increasing $H_0$ try to decrease $r_s$. This is the main idea behind the early modifications of $\Lambda$CDM model including. This effectively means the last scattering surface is closer to the big-bang. This distance is given by
 \begin{eqnarray}
r_s^\star=\int_{z_\star}^\infty\, c_s(z)\, \frac{1}{H(z)}\,dz,
\end{eqnarray}
 where $c_s$ is the speed of sound of the photon-baryon fluid and $z_\star$ is the redshift of the CMB last scattering. So what we can do alongside reducing $r_s$ is to play with the Hubble parameter $H(z)$ and the speed of sound $c_s(z)$ at early times (i.e. $z>z_\star$). For the case of modification on $c_s(z)$ there are many examples including \cite{Jedamzik:2020krr,Sekiguchi:2020teg}. The modification of $H(z)$ at early times is the main idea behind different approaches of early dark energy \cite{Kamionkowski:2022pkx}.

\section{The $H_0$ tension in the CCPot model}
In the CCPot model, the stiff matter era makes the early cosmological evolution modified. Consequently, it shows its effects on the sound horizon, $r^\star_s$. The effect is in the reduction of $r^\star_s$ as it could be guessed due to larger $H(z)$ for very high redshifts. This may give a chance to have higher $H_0$ values and address the Hubble tension. 

\subsection{Datasets}
In this work, we use the  CMB, BAO and local $H_0$ datasets to constrain our model. It is important to mention that for the CMB we have not used the full dataset but the reduced. It has been shown that the reduced CMB dataset captures the main information in the CMB and is useful to check the models beyond the $\Lambda$CDM \cite{Mukherjee:2008kd,Tutusaus:2023cms}. The reduced CMB dataset includes the angular scale of the sound horizon at the last scattering surface $\theta_\star$, the CMB shift parameter $R=\sqrt{\Omega_m H_0^2}D_M(z^\star)$ and the baryon density and the spectral index. Since our model does not affect the spectral index and the baryonic physics, we do not expect any modification in these two parameters and we fix them same as their best values from Planck \cite{Aghanim:2018eyx}. We also do not use the CMB shift parameter to constrain our model but will show that our final prediction for it is compatible with its value from Planck $R=1.7478\pm0.0046$ \cite{Tutusaus:2023cms}. So from the CMB, we use the acoustic angular angle $\theta_\star = 1.04090\pm0.00031 $ which is very well constrained  \cite{Aghanim:2018eyx}.  To have more information about the sound horizon, we also use the late time BAO observations. For our purposes, the BAO's are very crucial. Their main physics is exactly the CMB physics but they are far from the last scattering surface. This property is one of the reasons that makes late time solution to the Hubble tension very hard. Since any late time modification in the cosmological distances may affect the CMB distance (and consequently the $H_0$ deduced from CMB) but cannot change the distances given by the BAO's. Because of this reason we have used some independent BAO datasets: $r_d/D_V=0.336\pm0.015$ \cite{beutler20116df}; $D_A/(1.81r_d)=10.75\pm0.43$ \cite{DES:2017rfo}; $D_V/r_d=4.466\pm0.168$ \cite{Ross:2014qpa};  $D_V/r_d=11.548\pm0.559$, $D_V/r_d=14.946\pm0.68\pm0.559$, $D_V/r_d=16.931\pm0.579$ \cite{Kazin:2014qga};  $D_V/r_d=16.085\pm0.406$ \cite{Bautista:2017wwp}; $D_M/r_d=36.6\pm1.35$, $D_H/r_d=8.94\pm0.225$ \cite{duMasdesBourboux:2017mrl}; $D_H/r_d=9.07\pm.031$  \cite{Bautista:2017zgn}; $D_V/r_d=9.995\pm 0.108$, $D_V/r_d=12.701\pm0.129$, $D_V/r_d=14.481\pm0.149$ \cite{BOSS:2016wmc}; $r_d\,H=25500\pm1800$, $D_M/(2.25 r_d)=12.58\pm0.70$ \cite{Ata:2017dya}. Note that the measurements are at different redshifts. For the direct local measurement of the $H_0$ we use  $H_0 = 73.30\pm1.04$ \cite{Riess:2021jrx}. First we check the CCPot model's parameters against $\theta+BAO$ (i.e. reduced CMB and all the BAO's). Then if the model was compatible with higher values of $H_0$, we are allowed to add the local $H_0$ measurement.

\subsection{Results}

In the CCPot model, we have an additional $\beta$ parameter as it appeared in (\ref{wCCPot}). In addition to this parameter, we allow the dark matter density ($\Omega_{c}$) and the Hubble constant ($H_0$) to be free parameters. We do not assume the other parameters as baryon density ($\Omega_b$), spectral index ($n_s$) and etc. to be free parameters as we mentioned above. We also impose spatial flatness which results in 
\begin{eqnarray}\nonumber
H^2&=&H_0^2\bigg[ \Omega_r\,(1+z)^4 + \Omega_m\,(1+z)^3\\\nonumber&+&(1-\Omega_r-\Omega_m)\,\bigg((1+e^{-\beta} (1+z)^6)/(1+e^{-\beta})\bigg) \bigg]
\end{eqnarray}
as the Friedmann equation where $\Omega_r=\Omega_\gamma+\Omega_\nu$ and $\Omega_m=\Omega_b+\Omega_c$.

\begin{figure}
	\begin{center}
		\includegraphics[width=8.5cm]{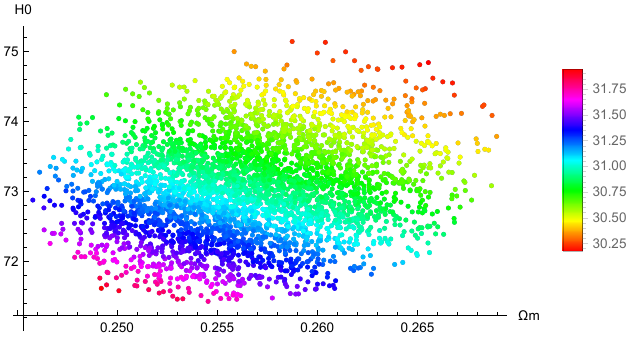}
		\caption{This plot shows the $1\sigma$ contour in the $H_0-\Omega_m$ plane while the color dots (bar) is representing the $\beta$ parameter. In this plot all the $\theta+BAO$ and  $H_0$ datasets are used. The results are totally compatible with higher $H_0$ values while has no conflict with the $\theta+BAO$. It is obvious that for a fixed $\beta$ value (a fixed color) the expected anti-correlation between $H_0$ and $\Omega_m$ is seen.} 
		\label{H0-Om}
	\end{center}
\end{figure}

\begin{figure}
	\begin{center}
		\includegraphics[width=8.5cm]{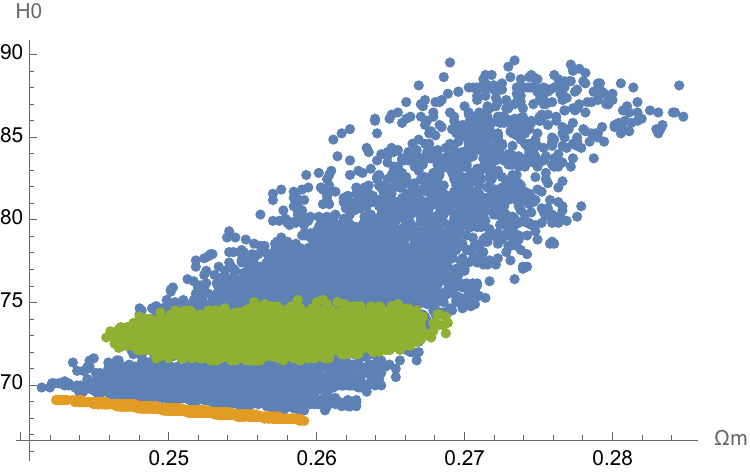}
		\caption{The blue region is $1\sigma$ likelihood for the CCPot model which is constrained by only $\theta+BAO$ dataset. Obviously, it is compatible with the higher local values of $H_0$ which allows us to constrain the CCPot model with $\theta+BAO+H_0$ dataset. For this case, the green region is the $1\sigma$ likelihood for the CCPot model. The orange region is the standard $\Lambda$CDM model which is constrained by $\theta+BAO$ that shows inconsistency with the local $H_0$ value. The  $\Lambda$CDM is compatible with the CCPot as we could expect (because it is the $\beta\rightarrow\infty$ limit of the CCPot). But in general the CCPot prefers higher values for the $H_0$. This means not also the variance but the mean of the $H_0$ increases. The green region, is plotted in figure \ref{H0-Om} in more details. } 
		\label{H0-Om-full}
	\end{center}
\end{figure}

The figure \ref{H0-Om} shows the $1\sigma$ posterior on our free parameters when the CCPot is constrained by all the datasets $\theta+BAO$ and late $H_0$ measurement. The result is very consistent with the higher value of the late $H_0$ measurements which means there is no Hubble tension in our model. However, it should be emphasized that before allowing to add datasets together we had to check their consistencies. In figure \ref{H0-Om-full} we have checked the CCPot against $\theta+BAO$ without $H_0$ dataset. The figure shows that the posterior for the CCPot is big enough to contain higher value of $H_0$. Note that it is not just because of larger variance but also is due to higher value for the mean value of $H_0$. The $\Lambda$CDM as a limit of the CCPot (for $\beta\rightarrow 0$) is at the border of $1\sigma$ region. Back to figure \ref{H0-Om}, we can see that for a fixed value of $\beta$, we can see the expected anti-correlation of $H_0$ and $\Omega_m$. The best values and their $1\sigma$ variances are reported in Table \ref{table}. As an independent check, we checked the CCPot results against the CMB shift parameter which is almost model independent \cite{Mukherjee:2008kd,Tutusaus:2023cms} as we already mentioned. The CCPot prediction (when we use the best fit parameters) is $R=1.7462$ which is in $1\sigma$ prediction by Planck results $R=1.7478\pm0.0046$.

\begin{figure}
	\begin{center}
		\includegraphics[width=8.5cm]{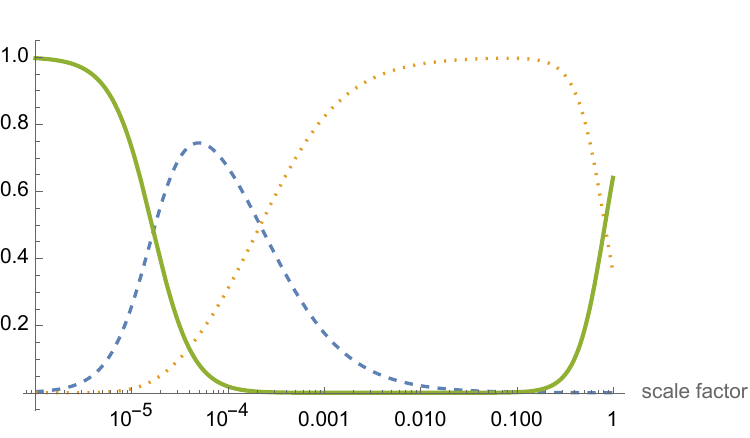}
		\caption{The solid line is the CCPot fluid time dependence in the history of the cosmos. At the late time it is the standard cosmological constant ($w=-1$) and becomes dominant. The same fluid is dominant in the very early universe due to its similarities to the stiff matter ($w=+1$). The figure is plotted for the best values of parameters reported in Table \ref{table}.} 
		\label{fraction}
	\end{center}
\end{figure}

\begin{figure}
	\begin{center}
		\includegraphics[width=8.5cm]{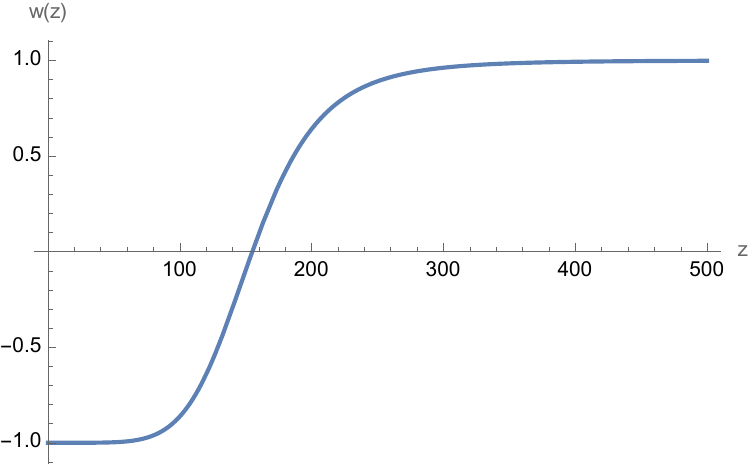}
		\caption{The (effective) equation of state for the CCPot fluid is plotted for the best fit parameters. The transition between the stiff matter ($w=+1$) and the cosmological constant ($w=-1$) is obvious.} 
		\label{wz}
	\end{center}
\end{figure}

\begin{table}
	\centering
\begin{tabular}{|c|c|c|c|c|c|c|c|c|c|}
	\hline
	$H_0$&$\Omega_m$&$\beta$\\
	\hline
$73.31^{+1.83}_{-1.88}$&$0.2578^{+0.111}_{-0.121}$&$30.80^{+1.13}_{-0.62}$\\
	\hline
\end{tabular}
\caption{The best fit values of our free parameters as well as their $1\sigma$ variances is reported. The parameters are constrained by using the full datasets: $\theta+BAO$ and $H_0$.}\label{table}
\end{table}

 In figure \ref{fraction}, the fraction of each energy content is plotted against redshift. As we expected the new fluid is dominant before $z\sim10^5$ and after $z\sim0.3$. The former phase is the one which makes the $r_s$ less than the standard scenario. This phase predicts higher values for the Hubble parameter in  redshifts above $10^4-10^5$. On the other hand since the interaction rates $\Gamma$ are coming from the independent physics of particles so we expect them to be the same as the standard cosmology. In conclusion, we expect the particles be out of the equilibrium sooner with respect to the standard $\Lambda$CDM. This does not true for the photon decoupling since the cosmological modifications due to the CCPot are very small at the decoupling redshift and so the last scattering surface distance from us should be the same as the standard one. However, we have to say that the physics before the last scattering surface in our model needs more consideration which remains for the future works. In figure \ref{wz}, we can see the effective equation of state $w(z)$. The transition redshift between $w=+1$ and $w=-1$ is around $z\sim150$ for the best fit parameters' values. It seems hard to make observational constraint in these redshifts.  To go further in this direction, we need to use full CMB dataset to look for more details of the CCPot model in early/intermediate/late times, which remains for the future.

\section{Concluding remarks}
The cosmological constant, which is a constant in the Lagrangian, can explain the late time acceleration perfectly. But by another view, this constant is the remnant of the field rolling on a constant potential. In this viewpoint at the early times, the field (which is assumed to be scalar) behaves as a stiff fluid with $w=+1$ while it stops rolling due to the Hubble friction and results in a fluid with $w=-1$. We studied this model and could show that it can resolve the Hubble tension by reducing the sound horizon at the early times. We checked our model against the (reduced) CMB as well as BAO and they are consistent with each other and the late time measurements of the Hubble constant. However for the future works we would like to check the full model against the full CMB dataset.

In the theoretical side, the cosmological constant potential can be seen as the leftover of the early inflation. The inflaton field is rolling down the potential to give the inflationary phase and at the end it reaches a constant potential. The initial conditions of the CCPot part is the final conditions of the inflationary phase. This scenario needs more consideration and can be interesting if it can be related to the PBH production at the end of inflation. This idea seems natural to come to mind since the PBH production can be efficient if there is an ultra slow roll regime which has effectively $w=+1$.

\vspace{1.0 cm}
\textit{Acknowledgments:}
I would like to thank  Shant Baghram, Abdolali Banihashemi and Sara Khatibi for useful discussions about this paper. I am also grateful to Abdolali Banihashemi for allowing me to use his MCMC code.

\bibliography{CCPref}

\begin{thebibliography}{31}
\expandafter\ifx\csname natexlab\endcsname\relax\def\natexlab#1{#1}\fi
\expandafter\ifx\csname bibnamefont\endcsname\relax
  \def\bibnamefont#1{#1}\fi
\expandafter\ifx\csname bibfnamefont\endcsname\relax
  \def\bibfnamefont#1{#1}\fi
\expandafter\ifx\csname citenamefont\endcsname\relax
  \def\citenamefont#1{#1}\fi
\expandafter\ifx\csname url\endcsname\relax
  \def\url#1{\texttt{#1}}\fi
\expandafter\ifx\csname urlprefix\endcsname\relax\def\urlprefix{URL }\fi
\providecommand{\bibinfo}[2]{#2}
\providecommand{\eprint}[2][]{\url{#2}}

\bibitem[{\citenamefont{Riess et~al.}(2016)}]{R16}
\bibinfo{author}{\bibfnamefont{A.~G.} \bibnamefont{Riess}}
  \bibnamefont{et~al.}, \bibinfo{journal}{Astrophys. J.}
  \textbf{\bibinfo{volume}{826}}, \bibinfo{pages}{56} (\bibinfo{year}{2016}),
  \eprint{1604.01424}.

\bibitem[{\citenamefont{Riess et~al.}(2019)\citenamefont{Riess, Casertano,
  Yuan, Macri, and Scolnic}}]{Riess:2019cxk}
\bibinfo{author}{\bibfnamefont{A.~G.} \bibnamefont{Riess}},
  \bibinfo{author}{\bibfnamefont{S.}~\bibnamefont{Casertano}},
  \bibinfo{author}{\bibfnamefont{W.}~\bibnamefont{Yuan}},
  \bibinfo{author}{\bibfnamefont{L.~M.} \bibnamefont{Macri}}, \bibnamefont{and}
  \bibinfo{author}{\bibfnamefont{D.}~\bibnamefont{Scolnic}},
  \bibinfo{journal}{Astrophys. J.} \textbf{\bibinfo{volume}{876}},
  \bibinfo{pages}{85} (\bibinfo{year}{2019}), \eprint{1903.07603}.

\bibitem[{\citenamefont{Riess et~al.}(2022)}]{Riess:2021jrx}
\bibinfo{author}{\bibfnamefont{A.~G.} \bibnamefont{Riess}}
  \bibnamefont{et~al.}, \bibinfo{journal}{Astrophys. J. Lett.}
  \textbf{\bibinfo{volume}{934}}, \bibinfo{pages}{L7} (\bibinfo{year}{2022}),
  \eprint{2112.04510}.

\bibitem[{\citenamefont{Aghanim et~al.}(2020)}]{Aghanim:2018eyx}
\bibinfo{author}{\bibfnamefont{N.}~\bibnamefont{Aghanim}} \bibnamefont{et~al.}
  (\bibinfo{collaboration}{Planck}), \bibinfo{journal}{Astron. Astrophys.}
  \textbf{\bibinfo{volume}{641}}, \bibinfo{pages}{A6} (\bibinfo{year}{2020}),
  \eprint{1807.06209}.

\bibitem[{\citenamefont{Heymans et~al.}(2021)}]{Heymans:2020gsg}
\bibinfo{author}{\bibfnamefont{C.}~\bibnamefont{Heymans}} \bibnamefont{et~al.},
  \bibinfo{journal}{Astron. Astrophys.} \textbf{\bibinfo{volume}{646}},
  \bibinfo{pages}{A140} (\bibinfo{year}{2021}), \eprint{2007.15632}.

\bibitem[{\citenamefont{Schwarz et~al.}(2016)\citenamefont{Schwarz, Copi,
  Huterer, and Starkman}}]{Schwarz:2015cma}
\bibinfo{author}{\bibfnamefont{D.~J.} \bibnamefont{Schwarz}},
  \bibinfo{author}{\bibfnamefont{C.~J.} \bibnamefont{Copi}},
  \bibinfo{author}{\bibfnamefont{D.}~\bibnamefont{Huterer}}, \bibnamefont{and}
  \bibinfo{author}{\bibfnamefont{G.~D.} \bibnamefont{Starkman}},
  \bibinfo{journal}{Class. Quant. Grav.} \textbf{\bibinfo{volume}{33}},
  \bibinfo{pages}{184001} (\bibinfo{year}{2016}), \eprint{1510.07929}.

\bibitem[{\citenamefont{Akrami et~al.}(2020)}]{Planck:2019evm}
\bibinfo{author}{\bibfnamefont{Y.}~\bibnamefont{Akrami}} \bibnamefont{et~al.}
  (\bibinfo{collaboration}{Planck}), \bibinfo{journal}{Astron. Astrophys.}
  \textbf{\bibinfo{volume}{641}}, \bibinfo{pages}{A7} (\bibinfo{year}{2020}),
  \eprint{1906.02552}.

\bibitem[{\citenamefont{Di~Valentino et~al.}(2021)\citenamefont{Di~Valentino,
  Mena, Pan, Visinelli, Yang, Melchiorri, Mota, Riess, and
  Silk}}]{DiValentino:2021izs}
\bibinfo{author}{\bibfnamefont{E.}~\bibnamefont{Di~Valentino}},
  \bibinfo{author}{\bibfnamefont{O.}~\bibnamefont{Mena}},
  \bibinfo{author}{\bibfnamefont{S.}~\bibnamefont{Pan}},
  \bibinfo{author}{\bibfnamefont{L.}~\bibnamefont{Visinelli}},
  \bibinfo{author}{\bibfnamefont{W.}~\bibnamefont{Yang}},
  \bibinfo{author}{\bibfnamefont{A.}~\bibnamefont{Melchiorri}},
  \bibinfo{author}{\bibfnamefont{D.~F.} \bibnamefont{Mota}},
  \bibinfo{author}{\bibfnamefont{A.~G.} \bibnamefont{Riess}}, \bibnamefont{and}
  \bibinfo{author}{\bibfnamefont{J.}~\bibnamefont{Silk}},
  \bibinfo{journal}{Class. Quant. Grav.} \textbf{\bibinfo{volume}{38}},
  \bibinfo{pages}{153001} (\bibinfo{year}{2021}), \eprint{2103.01183}.

\bibitem[{\citenamefont{Efstathiou}(2021)}]{Efstathiou:2021ocp}
\bibinfo{author}{\bibfnamefont{G.}~\bibnamefont{Efstathiou}},
  \bibinfo{journal}{Mon. Not. Roy. Astron. Soc.}
  \textbf{\bibinfo{volume}{505}}, \bibinfo{pages}{3866} (\bibinfo{year}{2021}),
  \eprint{2103.08723}.

\bibitem[{\citenamefont{Kamionkowski and Riess}(2023)}]{Kamionkowski:2022pkx}
\bibinfo{author}{\bibfnamefont{M.}~\bibnamefont{Kamionkowski}}
  \bibnamefont{and} \bibinfo{author}{\bibfnamefont{A.~G.} \bibnamefont{Riess}},
  \bibinfo{journal}{Ann. Rev. Nucl. Part. Sci.} \textbf{\bibinfo{volume}{73}},
  \bibinfo{pages}{153} (\bibinfo{year}{2023}), \eprint{2211.04492}.

\bibitem[{\citenamefont{Knox and Millea}(2020)}]{Knox:2019rjx}
\bibinfo{author}{\bibfnamefont{L.}~\bibnamefont{Knox}} \bibnamefont{and}
  \bibinfo{author}{\bibfnamefont{M.}~\bibnamefont{Millea}},
  \bibinfo{journal}{Phys. Rev. D} \textbf{\bibinfo{volume}{101}},
  \bibinfo{pages}{043533} (\bibinfo{year}{2020}), \eprint{1908.03663}.

\bibitem[{\citenamefont{Poulin et~al.}(2019)\citenamefont{Poulin, Smith,
  Karwal, and Kamionkowski}}]{Poulin:2018cxd}
\bibinfo{author}{\bibfnamefont{V.}~\bibnamefont{Poulin}},
  \bibinfo{author}{\bibfnamefont{T.~L.} \bibnamefont{Smith}},
  \bibinfo{author}{\bibfnamefont{T.}~\bibnamefont{Karwal}}, \bibnamefont{and}
  \bibinfo{author}{\bibfnamefont{M.}~\bibnamefont{Kamionkowski}},
  \bibinfo{journal}{Phys. Rev. Lett.} \textbf{\bibinfo{volume}{122}},
  \bibinfo{pages}{221301} (\bibinfo{year}{2019}), \eprint{1811.04083}.

\bibitem[{\citenamefont{Vagnozzi}(2023)}]{Vagnozzi:2023nrq}
\bibinfo{author}{\bibfnamefont{S.}~\bibnamefont{Vagnozzi}},
  \bibinfo{journal}{Universe} \textbf{\bibinfo{volume}{9}},
  \bibinfo{pages}{393} (\bibinfo{year}{2023}), \eprint{2308.16628}.

\bibitem[{\citenamefont{Ellis et~al.}(2007)\citenamefont{Ellis, Maartens, and
  MacCallum}}]{Ellis:2007ic}
\bibinfo{author}{\bibfnamefont{G.}~\bibnamefont{Ellis}},
  \bibinfo{author}{\bibfnamefont{R.}~\bibnamefont{Maartens}}, \bibnamefont{and}
  \bibinfo{author}{\bibfnamefont{M.~A.~H.} \bibnamefont{MacCallum}},
  \bibinfo{journal}{Gen. Rel. Grav.} \textbf{\bibinfo{volume}{39}},
  \bibinfo{pages}{1651} (\bibinfo{year}{2007}), \eprint{gr-qc/0703121}.

\bibitem[{\citenamefont{Chavanis}(2015)}]{Chavanis:2014lra}
\bibinfo{author}{\bibfnamefont{P.-H.} \bibnamefont{Chavanis}},
  \bibinfo{journal}{Phys. Rev. D} \textbf{\bibinfo{volume}{92}},
  \bibinfo{pages}{103004} (\bibinfo{year}{2015}), \eprint{1412.0743}.

\bibitem[{\citenamefont{Zel'dovich}(1961)}]{Zeldovich:1961sbr}
\bibinfo{author}{\bibfnamefont{Y.~B.} \bibnamefont{Zel'dovich}},
  \bibinfo{journal}{Zh. Eksp. Teor. Fiz.} \textbf{\bibinfo{volume}{41}},
  \bibinfo{pages}{1609} (\bibinfo{year}{1961}).

\bibitem[{\citenamefont{Dimopoulos}(2017)}]{Dimopoulos:2017ged}
\bibinfo{author}{\bibfnamefont{K.}~\bibnamefont{Dimopoulos}},
  \bibinfo{journal}{Phys. Lett. B} \textbf{\bibinfo{volume}{775}},
  \bibinfo{pages}{262} (\bibinfo{year}{2017}), \eprint{1707.05644}.

\bibitem[{\citenamefont{Aylor et~al.}(2019)\citenamefont{Aylor, Joy, Knox,
  Millea, Raghunathan, and Wu}}]{Aylor:2018drw}
\bibinfo{author}{\bibfnamefont{K.}~\bibnamefont{Aylor}},
  \bibinfo{author}{\bibfnamefont{M.}~\bibnamefont{Joy}},
  \bibinfo{author}{\bibfnamefont{L.}~\bibnamefont{Knox}},
  \bibinfo{author}{\bibfnamefont{M.}~\bibnamefont{Millea}},
  \bibinfo{author}{\bibfnamefont{S.}~\bibnamefont{Raghunathan}},
  \bibnamefont{and} \bibinfo{author}{\bibfnamefont{W.~L.~K.} \bibnamefont{Wu}},
  \bibinfo{journal}{Astrophys. J.} \textbf{\bibinfo{volume}{874}},
  \bibinfo{pages}{4} (\bibinfo{year}{2019}), \eprint{1811.00537}.

\bibitem[{\citenamefont{Jedamzik and Pogosian}(2020)}]{Jedamzik:2020krr}
\bibinfo{author}{\bibfnamefont{K.}~\bibnamefont{Jedamzik}} \bibnamefont{and}
  \bibinfo{author}{\bibfnamefont{L.}~\bibnamefont{Pogosian}},
  \bibinfo{journal}{Phys. Rev. Lett.} \textbf{\bibinfo{volume}{125}},
  \bibinfo{pages}{181302} (\bibinfo{year}{2020}), \eprint{2004.09487}.

\bibitem[{\citenamefont{Sekiguchi and Takahashi}(2021)}]{Sekiguchi:2020teg}
\bibinfo{author}{\bibfnamefont{T.}~\bibnamefont{Sekiguchi}} \bibnamefont{and}
  \bibinfo{author}{\bibfnamefont{T.}~\bibnamefont{Takahashi}},
  \bibinfo{journal}{Phys. Rev. D} \textbf{\bibinfo{volume}{103}},
  \bibinfo{pages}{083507} (\bibinfo{year}{2021}), \eprint{2007.03381}.

\bibitem[{\citenamefont{Mukherjee et~al.}(2008)\citenamefont{Mukherjee, Kunz,
  Parkinson, and Wang}}]{Mukherjee:2008kd}
\bibinfo{author}{\bibfnamefont{P.}~\bibnamefont{Mukherjee}},
  \bibinfo{author}{\bibfnamefont{M.}~\bibnamefont{Kunz}},
  \bibinfo{author}{\bibfnamefont{D.}~\bibnamefont{Parkinson}},
  \bibnamefont{and} \bibinfo{author}{\bibfnamefont{Y.}~\bibnamefont{Wang}},
  \bibinfo{journal}{Phys. Rev. D} \textbf{\bibinfo{volume}{78}},
  \bibinfo{pages}{083529} (\bibinfo{year}{2008}), \eprint{0803.1616}.

\bibitem[{\citenamefont{Tutusaus et~al.}(2023)\citenamefont{Tutusaus, Kunz, and
  Favre}}]{Tutusaus:2023cms}
\bibinfo{author}{\bibfnamefont{I.}~\bibnamefont{Tutusaus}},
  \bibinfo{author}{\bibfnamefont{M.}~\bibnamefont{Kunz}}, \bibnamefont{and}
  \bibinfo{author}{\bibfnamefont{L.}~\bibnamefont{Favre}}
  (\bibinfo{year}{2023}), \eprint{2311.16862}.

\bibitem[{\citenamefont{Beutler et~al.}(2011)\citenamefont{Beutler, Blake,
  Colless, Jones, Staveley-Smith, Campbell, Parker, Saunders, and
  Watson}}]{beutler20116df}
\bibinfo{author}{\bibfnamefont{F.}~\bibnamefont{Beutler}},
  \bibinfo{author}{\bibfnamefont{C.}~\bibnamefont{Blake}},
  \bibinfo{author}{\bibfnamefont{M.}~\bibnamefont{Colless}},
  \bibinfo{author}{\bibfnamefont{D.~H.} \bibnamefont{Jones}},
  \bibinfo{author}{\bibfnamefont{L.}~\bibnamefont{Staveley-Smith}},
  \bibinfo{author}{\bibfnamefont{L.}~\bibnamefont{Campbell}},
  \bibinfo{author}{\bibfnamefont{Q.}~\bibnamefont{Parker}},
  \bibinfo{author}{\bibfnamefont{W.}~\bibnamefont{Saunders}}, \bibnamefont{and}
  \bibinfo{author}{\bibfnamefont{F.}~\bibnamefont{Watson}},
  \bibinfo{journal}{Monthly Notices of the Royal Astronomical Society}
  \textbf{\bibinfo{volume}{416}}, \bibinfo{pages}{3017} (\bibinfo{year}{2011}).

\bibitem[{\citenamefont{Abbott et~al.}(2019)}]{DES:2017rfo}
\bibinfo{author}{\bibfnamefont{T.~M.~C.} \bibnamefont{Abbott}}
  \bibnamefont{et~al.} (\bibinfo{collaboration}{DES}), \bibinfo{journal}{Mon.
  Not. Roy. Astron. Soc.} \textbf{\bibinfo{volume}{483}}, \bibinfo{pages}{4866}
  (\bibinfo{year}{2019}), \eprint{1712.06209}.

\bibitem[{\citenamefont{Ross et~al.}(2015)\citenamefont{Ross, Samushia,
  Howlett, Percival, Burden, and Manera}}]{Ross:2014qpa}
\bibinfo{author}{\bibfnamefont{A.~J.} \bibnamefont{Ross}},
  \bibinfo{author}{\bibfnamefont{L.}~\bibnamefont{Samushia}},
  \bibinfo{author}{\bibfnamefont{C.}~\bibnamefont{Howlett}},
  \bibinfo{author}{\bibfnamefont{W.~J.} \bibnamefont{Percival}},
  \bibinfo{author}{\bibfnamefont{A.}~\bibnamefont{Burden}}, \bibnamefont{and}
  \bibinfo{author}{\bibfnamefont{M.}~\bibnamefont{Manera}},
  \bibinfo{journal}{Mon. Not. Roy. Astron. Soc.}
  \textbf{\bibinfo{volume}{449}}, \bibinfo{pages}{835} (\bibinfo{year}{2015}),
  \eprint{1409.3242}.

\bibitem[{\citenamefont{Kazin et~al.}(2014)}]{Kazin:2014qga}
\bibinfo{author}{\bibfnamefont{E.~A.} \bibnamefont{Kazin}}
  \bibnamefont{et~al.}, \bibinfo{journal}{Mon. Not. Roy. Astron. Soc.}
  \textbf{\bibinfo{volume}{441}}, \bibinfo{pages}{3524} (\bibinfo{year}{2014}),
  \eprint{1401.0358}.

\bibitem[{\citenamefont{Bautista et~al.}(2018)}]{Bautista:2017wwp}
\bibinfo{author}{\bibfnamefont{J.~E.} \bibnamefont{Bautista}}
  \bibnamefont{et~al.}, \bibinfo{journal}{Astrophys. J.}
  \textbf{\bibinfo{volume}{863}}, \bibinfo{pages}{110} (\bibinfo{year}{2018}),
  \eprint{1712.08064}.

\bibitem[{\citenamefont{du~Mas~des Bourboux
  et~al.}(2017)}]{duMasdesBourboux:2017mrl}
\bibinfo{author}{\bibfnamefont{H.}~\bibnamefont{du~Mas~des Bourboux}}
  \bibnamefont{et~al.}, \bibinfo{journal}{Astron. Astrophys.}
  \textbf{\bibinfo{volume}{608}}, \bibinfo{pages}{A130} (\bibinfo{year}{2017}),
  \eprint{1708.02225}.

\bibitem[{\citenamefont{Bautista et~al.}(2017)}]{Bautista:2017zgn}
\bibinfo{author}{\bibfnamefont{J.~E.} \bibnamefont{Bautista}}
  \bibnamefont{et~al.}, \bibinfo{journal}{Astron. Astrophys.}
  \textbf{\bibinfo{volume}{603}}, \bibinfo{pages}{A12} (\bibinfo{year}{2017}),
  \eprint{1702.00176}.

\bibitem[{\citenamefont{Alam et~al.}(2017)}]{BOSS:2016wmc}
\bibinfo{author}{\bibfnamefont{S.}~\bibnamefont{Alam}} \bibnamefont{et~al.}
  (\bibinfo{collaboration}{BOSS}), \bibinfo{journal}{Mon. Not. Roy. Astron.
  Soc.} \textbf{\bibinfo{volume}{470}}, \bibinfo{pages}{2617}
  (\bibinfo{year}{2017}), \eprint{1607.03155}.

\bibitem[{\citenamefont{Ata et~al.}(2018)}]{Ata:2017dya}
\bibinfo{author}{\bibfnamefont{M.}~\bibnamefont{Ata}} \bibnamefont{et~al.},
  \bibinfo{journal}{Mon. Not. Roy. Astron. Soc.}
  \textbf{\bibinfo{volume}{473}}, \bibinfo{pages}{4773} (\bibinfo{year}{2018}),
  \eprint{1705.06373}.

\end{thebibliography}

	


\end{document}